\documentclass[12pt]{article}
\usepackage{amssymb}
\usepackage{amsbsy}

\def\k{{\boldsymbol{k}}}
\def\p{{\boldsymbol{p}}}
\def\x{{\boldsymbol{x}}} 
\def\f{{\boldsymbol{\mathcal{F}}}}

\begin{document}

\title{\bf Momentum Dependent Fundamental Action}

\author{\bf  Karl De Paepe}

\maketitle

\begin{abstract} 
\noindent   The fundamental action is dependent on momentum. Some consequences are 
presented for matter waves and scalar quantum field theory.  
\end{abstract}

  
\section{Introduction}

This article is a revision of K. De Paepe~\cite{kd}. We present an argument as to why the fundamental 
action or Planck's ``constant'' is dependent 
on momentum. Consider an unbound  particle with wave function $\psi(\x,t)$ with compact support. 
We then have by
\begin{equation}\label{E:p}
\phi(\p)= \int \frac{d^3 \x}{(2 \pi)^{3/2}}\psi(\x, 0) e^{-i \frac{\p \cdot \x}{\hbar}} 
\end{equation}
that $\phi(\p)$ is an entire function. There is then a nonzero probability the particle has $|\p|$ larger
than any given value and hence for an unbound particle an energy 
larger than that of the universe which is impossible. The de Broglie relation is
\begin{equation}\label{de}
\p =\hbar \k.
\end{equation}
Instead of Eq.~\ref{E:p} use the following equation
\begin{eqnarray}\label{f}
\varphi(\k)=\int \frac{d^3 \x}{(2 \pi)^{3/2}}\psi(\x, 0) e^{-i \k \cdot \x}.
\end{eqnarray}
Eq.~\ref{f} is just the inverse Fourier transform of the wave function 
whereas Eq.~\ref{E:p} involves the  inverse Fourier transform and the 
assumption of Eq.~\ref{de}. The fact that  $\p \rightarrow \infty $ as $\k \rightarrow \infty$ as
happens in Eq.~\ref{de} leads to the impossibility as presented above.
We must have instead that 
\begin{eqnarray}\label{E: bar}
\p=\hbar(\p) \k, \quad  |\p| \le p_{max}
\end{eqnarray}
for all $\k$ for some finite maximum magnitude of momentum $p_{max}$. 
This implies that  $ \hbar(\p) \rightarrow 0$ as $\k \rightarrow \infty$. We could be more general and have
instead the fundamental action dependent on $\k$ but in this revised article we will restrict 
to a fundamental action dependent on momentum and a momentum that is an increasing function of $|\k|$. 
By Eq.~\ref{f}
there is a nonzero probability that the particle with wave function as previous has  
a  $|\k|$ larger than any given value. The magnitude of the momentum will however by 
Eq.~\ref{E: bar} always be less than or equal to $p_{max}$. 

We require that Eq.~\ref{E: bar} 
hold for ``simple'' particles. Define $\hbar(\p)$ to be zero for $|\p|$ larger than $p_{max}$. 
Now $\hbar(\p)$ could instead depend on energy and $E_{max}$ would the maximum energy. A macroscopic
object with energy larger than $E_{max}$ could have zero 
de Broglie wavelength and so behave as a classical particle. This could hold for a 
measuring apparatus providing a possible resolution of the quantum measurement paradox. 
A measurement results in a quantum system interacting with a classical measuring 
apparatus resulting in a classical system not a superposition of quantum states. 
  
We choose units so that the fundamental action is equal to $2 \pi$ 
and the speed of light is equal to one. We choose for Minkowski metric 
\begin{eqnarray}
\eta_{11}= \eta_{22}= \eta_{33}= - \eta_{44}=1, \quad \eta_{\mu \nu}=0 \quad \mbox{if} \quad \mu \not= \nu.
\end{eqnarray}
In what follows we will require throughout Lorentz invariance. In the following sections we present 
some interesting consequences of a momentum dependent fundamental action for matter waves and 
scalar quantum field theory. 

\section{Matter Waves}

For the matter wave of a particle with fundamental action dependent on momentum
\begin{eqnarray}\label{ma}
\p = \hbar(\p) \k, \quad \hspace{.2 cm} E= \hbar (\p) \omega.
\end{eqnarray}
Now $(\k, \omega)$ is a four-vector and for a free particle 
$(\p,E)$ is a four-vector so Eqs.~\ref{ma} imply that  $\hbar(\p)$ must be  Lorentz 
invariant and hence $\hbar(\p)$ depends only on $|\p|$. 

Consider a particle in a nonzero constant potential $V$. The energy of the particle is
\begin{equation}\label{re}
E=\sqrt{\p^2 +m^2} +V.
\end{equation}
We then have for a plane wave by Eqs.~\ref{ma} and Eq.~\ref{re} that
\begin{equation}\label{ev}
\omega^2-\k^2=2V\sqrt{\k^2+m^2\hbar^{-2}(\p)} +(V^2+m^2)\hbar^{-2}(\p).
\end{equation}
Now phase $\k \cdot \x - \omega t$ and hence $\omega^2-\k^2$ are Lorentz invariant so the right hand side
of Eq.~\ref{ev} must be Lorentz invariant. Without introducing a special inertial frame 
the right hand side of Eq.~\ref{ev} can not be made Lorentz invariant even when $\hbar(\p)=1$.

The inertial frame $\f$ that is at rest with respect to the distant stars is the only special inertial
frame. This inertial frame is needed in order to define acceleration and hence mass of any particle. 
In inertial frame $\f$ we can define basis vectors
\begin{eqnarray}
u_1=(1,0,0,0), u_2=(0,1,0,0), u_3=(0,0,1,0), u_4=(0,0,0,1)
\end{eqnarray} 
and four-vector
\begin{eqnarray}
p=(\p, \sqrt{\p^2 + m^2}).
\end{eqnarray}
Since
\begin{eqnarray}
u_4 \cdot p = \sqrt{(u_1 \cdot p)^2 +(u_2 \cdot p)^2 +(u_3 \cdot p)^2 +m^2} 
\end{eqnarray}
we have that any Lorentz invariant function constructed from vectors $p,u_\mu$
which in the frame $\f$ is a function of $|\p|$ must be equal to a function of 
\begin{eqnarray}
L(p)= \sqrt{(u_1 \cdot p)^2 +(u_2 \cdot p)^2 +(u_3 \cdot p)^2}.
\end{eqnarray}
Define $\hbar(p)$ to be the function of $L(p)$ that in the inertial frame 
$\f$ is the function $\hbar(\p)$. 

For
\begin{eqnarray}\label{lm}
\hbar(\p)=\left\{\begin{array}{ll}
\sqrt{M^2-\p^2}  &  |\p|\le M \\
0 & |\p| > M \end{array} \right.
\end{eqnarray}
where $M$ is a very large mass we have using using Eq.~\ref{ev} and Eq.~\ref{lm} with $V=0$ 
for group velocity of the matter wave
\begin{equation}\label{na}
\nabla_{\k} \omega=\frac{(1+ \frac{m^2}{M^2})\k}{\sqrt{(1+\frac{m^2}{M^2})\k^2+m^2}}.
\end{equation}
The particle velocity ${\boldsymbol{v}}$ is using Eqs.~\ref{ma} and Eq.~\ref{na}
\begin{eqnarray}
{\boldsymbol{v}} = \frac{ \p}{E} = \frac{\k}{\omega} = \frac{1}{1+\frac{m^2}{M^2}}\nabla_\k \omega.
\end{eqnarray}
By Eq.~\ref{re} with $V=0$ we have  $|{\boldsymbol{v}}| <1 $.
As $\k \rightarrow \infty$
\begin{equation}
|\nabla_{\k} \omega| \rightarrow \sqrt{1+\frac{m^2}{M^2}}.
\end{equation}
The magnitude of group velocity for large $\k$ will be slightly larger than the speed of light. 
If a macroscopic object with mass much larger than $M$ could be made into a ``simple'' particle 
its magnitude of group velocity could be much larger than the speed of light. 

\section{ Scalar Quantum Field Theory}

Let the  self-adjoint operator $\phi(x)$ represent  the scalar quantum field. We can define operators 
$q(\p, t)$ and $ p(\p, t)$ by  
\begin{eqnarray}\label{phi}
\phi(x)=\int \frac{d^3 \p} {(2 \pi)^{3/2}} q(\p, t) e^{i \p \cdot \x}, \quad p(\p, t) = 
\dot{q}(- \p, t)    
\end{eqnarray} 
where $q(\p, t)$ satisfies
\begin{eqnarray}\label{qcom}
[ q(\p, t) , q( \p' ,t )]= 0 
\end{eqnarray} 
and 
\begin{eqnarray}\label{pcom}
[ q(\p, t) , p( \p' ,t )]= i \hbar( \p)  \delta^{(3)} ( \p - \p'). 
\end{eqnarray} 
By  Eq.~\ref{phi} and Eq.~\ref{pcom}, 
\begin{eqnarray}\label{dot}
[\phi(\x', t), \partial_4 \phi(\x, t)]= i \int \frac{d^3 \p}{(2 \pi)^3}  
e^{i \p \cdot (\x' -\x)} \hbar(\p).
\end{eqnarray} 
Since $\hbar(\p)$ has compact support the right hand side of Eq.~\ref{dot} is an entire function in 
$\x'-\x$. There is then a point $\x$ such that  
\begin{eqnarray}\label{partial}
[\phi(0),  \partial_4\phi(\x, 0)] \not=0.
\end{eqnarray}
Eq.~\ref{partial}  implies there must be value of $t$ 
such that  $\x^2 - t^2  >0$ and 
\begin{eqnarray}\label{momen}
[ \phi( 0),  \phi(\x, t)]  \not=0.
\end{eqnarray} 
This is however zero for the  scalar quantum field theory where 
$\hbar(\p)=1$. Eq.~\ref{momen} implies there are measurements that are separated by a space-like 
interval that can influence each other.  

Consider the following free scalar field
\begin{eqnarray}
\phi(x) =\int \frac{d^3 \p}{(2 \pi)^{3/2}} \frac{\hbar^{1/2}(\p)}
{(2 p^4)^{1/2}} (e^{i p \cdot x } a(\p) + e^{-i p \cdot x} a^\dagger(\p))
\end{eqnarray}where
\begin{eqnarray}
p^4=\sqrt{\p^2 +m^2}, \quad [a(\p), a^\dagger(\p') ]=  \delta^{(3)}(\p-\p'), 
\quad  [a(\p), a(\p') ]= 0.
\end{eqnarray}
Comparing with Eq.~\ref{phi} we have
\begin{eqnarray}
q(\p , t)= \frac{\hbar^{1/2}(\p)}{(2 p^4)^{1/2}} (e^{-i p^4 t} a(\p) + e^{i p^4 t} a^\dagger(-\p)).
\end{eqnarray}
Now $q(\p, t)$ satisfies Eq.~\ref{qcom}, Eq.~\ref{pcom} and
\begin{eqnarray}
i \hbar(\p) {\dot{q}}(\p, t) = [q(\p, t), \frac{1}{2} \int d^3 \x \Big\{ {\dot \phi}^2+ (\nabla \phi)^2
+m^2 \phi^2 \Big\}].
\end{eqnarray}
Since $\hbar(\p)$ is zero for $|\p|>p_{max}$ the total zero-point energy
\begin{eqnarray}
\frac{1}{2} \int d^3 \p \hbar(\p) \sqrt{\p^2 + m^2}
\end{eqnarray}
will be finite. A finite zero-point energy would be required in a well founded theory of general 
relativity for a scalar field. The free scalar field satisfies
\begin{eqnarray}\label{E: modify}
[\phi(x), \phi(x')]=  \int \frac{d^3 \p}{ (2 \pi)^3 2 p^4}\hbar(p)
(e^{i p \cdot ( x-x')  }-  e^{-i p \cdot (x-x')}).
 \end{eqnarray}
It satisfies the Klein-Gordon equation and
\begin{eqnarray}
i \partial_4 \phi(x) =  [ \phi(x), H_0]
\end{eqnarray}
where  
\begin{eqnarray}\label{h0}
H_0= \int d^3 \p \sqrt{{\p}^2 + m^2} a^\dagger(\p) a(\p).
\end{eqnarray}

In the interaction picture with $H_0$ as in 
Eq.~\ref{h0}  the $\phi^3(x)$  interaction  one-loop diagram 
integral can be calculated to be 
\begin{eqnarray}
\frac{i \hbar(q)}{2} \int \frac{d^4 p}{(2 \pi)^4} \frac{\hbar(p)}{(p^2 + m^2- i \epsilon)} 
\frac{\hbar(p-q)}{((p-q)^2 + m^2 - i \epsilon)}.
\end{eqnarray} 
This diagram integral and in fact all S-matrix diagram integrals are finite and Lorentz invariant.

\vspace{.2cm}


\noindent \footnotesize{k.depaepe@utoronto.ca}

\end{document}